\documentclass[twocolumn,english]{revtex4-2}
\usepackage{mathpazo}
\usepackage[T1]{fontenc}
\usepackage{geometry}
\usepackage{color}
\usepackage{amsmath}
\usepackage{graphicx}

\makeatletter
\usepackage{amsmath,amsthm,amssymb}
\usepackage{graphicx,epstopdf}
\usepackage{color,bm,babel}
\usepackage[unicode=true,pdfusetitle,
 bookmarks=true,bookmarksnumbered=false,bookmarksopen=false,
 citecolor=cyan,urlcolor=magenta,linkcolor=blue, breaklinks=false,
 pdfborder={0 0 0},backref=false,colorlinks=true]{hyperref}
\usepackage{tcolorbox}

\usepackage{siunitx}
\makeatletter

\newcommand\figref[2]{Fig.\,\ref{#1}\hyperref[#1]{{#2}}}
\newcommand\figsref[2]{Figs.\,\ref{#1}\hyperref[#1]{{#2}}}
\newcommand\fref[2]{\ref{#1}\hyperref[#1]{{#2}}}

\makeatother

\begin{document}
\title{Thermal atoms facilitate intensity clipping between
vectorial dual-beam generated by a single metasurface chip}
\author{Chen Qing, Jialong Cui, Lishuang Feng, and Dengke Zhang}
\affiliation{School of Instrumentation and Optoelectronic Engineering, Beihang
University, Beijing 100191, China}
\address{Email: dkzhang@buaa.edu.cn}
\begin{abstract}
Manipulating vector beams is pivotal in fields such as particle manipulation,
image processing, and quantum communication. Flexibly adjusting the
intensity distribution of these beams is crucial for effectively realizing
these applications. This study introduces a vectorial dual-beam system
utilizing thermal atoms as the medium for modulating the intensity
profile of vector beams. A single metasurface is employed to generate
both the control and signal vector beams, each with unique vectorial
characteristics. The shaping of the signal beam profile is facilitated
by the interaction with thermal atoms, which can be controlled by
adjusting the control vector beam. This spatially selective absorption
is a result of the thermal atoms' response to the varying polarizations
within the vector beams. In this experiment, two distinct metasurface
chips are fabricated to generate vector beams with doughnut-shaped
and Gaussian-shaped intensity profiles. By adjusting the incident
power and polarization state of the control light, the doughnut-shaped
signal beams can be converted into a rotational dual-lobed pattern
or the dimensions of the Gaussian-distributed signal beams can be
modified. This study introduces a novel vector beam shaping technique
by integrating metasurfaces with thermal atoms, offering significant
promise for future applications requiring miniaturization, dynamic
operation, and versatile control capabilities.
\end{abstract}
\maketitle

\section{Introduction}

Vector beams feature an inhomogeneous polarization distribution across
their cross-section, offering enhanced flexibility for the spatial
manipulation and control of light fields \cite{Pattanayak1980,Hall1996}.
This characteristic enables their widespread use in various applications,
including optical manipulations \cite{matailoring,Kumar2024,donato2012optical,kozawa2010optical},
optical communications \cite{Milione2015,Zhu2021,zhao2015high},
and sensing technologies \cite{Syubaev2019,wu2021cylindrical}. Vector
beams are conventionally generated through the use of wave plates,
spatial light modulators, or digital micromirror devices. Although
these approaches facilitate the manipulation of light fields, they
also impose certain constraints and augment system complexity. In
recent years, metasurface technology has made considerable progress,
greatly improving the ability to manipulate light, especially in the
field of shaping vector beams. Advanced micro-nano fabrication techniques
have enabled the creation of nanoscale meta-atoms in metasurfaces,
thereby allowing for precise manipulation of the polarization and
intensity distributions of vector beams \cite{Guo2017,wen2022broadband,zhang2018optical,fu2023metasurface,cui2024multifunctional}.
The application of metasurfaces in vector beam manipulation presents
significant potential for advancements in optical tweezers \cite{zhu2022multidimensional}
and image processing \cite{Kim2021,arbabi2015dielectric,deng2018diatomic}.
To facilitate the dynamic modulation of vector beams and enhance their
manipulation, thereby enabling the investigation of diverse physical
phenomena and broadening their potential applications, it is imperative
to implement innovative control mechanisms. Recent advancements have
demonstrated that exploiting light-atom interactions presents a promising
approach for the control of vector beams.

In the context of light-atom interactions, the state of atoms is altered
by light, while simultaneously, atoms modify their properties in response
to the light. This process involves transitions between different
atomic energy levels through the absorption or emission of photons,
with the probability of these transitions being governed by selection
rules. Generally, the atomic response exhibits heightened sensitivity
to the polarization state of the incident light. Variations in polarization
states result in distinct atomic responses, thereby imparting a pronounced
polarization dependence to the absorption process. Consequently, the
utilization of atomic responses to vector beams has garnered extensive
investigation across various domains, including magnetic field sensing
\cite{Cai2024,Castellucci2021,Qiu2021}, optical spatial mode extraction
\cite{Chang2023} and optical storage \cite{Ye2019}. Nevertheless,
in the present scenario of vector beam-atom interactions, vector beams
are predominantly produced with traditional optical elements, which
restricts the diversity of vector beams configurations that can be
achieved.

Additionally, directly manipulating input signal parameters to control
vector beams may disrupt the stability of the optical path. Therefore,
incorporating additional physical fields to modulate the signal light
proves to be significantly more effective. This modulation can be
achieved by interacting with atoms using magnetic fields \cite{Yang2019,Sun2023,wang2024measuring},
electric fields \cite{Sedlacek2012a}, or light fields \cite{Hu2021}.
Among various methodologies, the use of a light field to shape another
light field is particularly noteworthy due to its exceptional efficiency,
simplicity, and minimal noise. This approach offers significant advantages
over the generation of magnetic and electrical fields, as well as
the complexities associated with assembling experimental setups.

In this paper, we propose a metasurface capable of generating vector
beams that exhibit a variety of polarization states along their beam
profiles. A vectorial dual-beam configuration has been developed,
consisting of control and signal beams that propagate concurrently
in the same direction through the metasurface chip. By incorporating
a thermal atomic vapor cell into the setup, a system is established
where the vectorial dual-beam engages with an atomic ensemble, allowing
for precise manipulation of the signal beam's intensity distribution.
In this arrangement, the incident signal light remains unchanged,
and the output signal beam's intensity profile is modulated by varying
the power or polarization state of the control light. This facilitates
a system where the signal beam is shaped or tailored by the control
beam. In the experiment, we designed and fabricated two metasurface
chips capable of generating vector beams with distinct polarizations
and intensity distributions. These chips were incorporated into the
system to demonstrate its capability in regulating various vector
beams configurations. Our approach capitalizes on the distinctive
sensitivity of thermal atoms to vector beams, enabling flexible, low-noise,
and highly efficient manipulation of the vector beams. 

\begin{figure*}
\begin{centering}
\includegraphics[width=14cm]{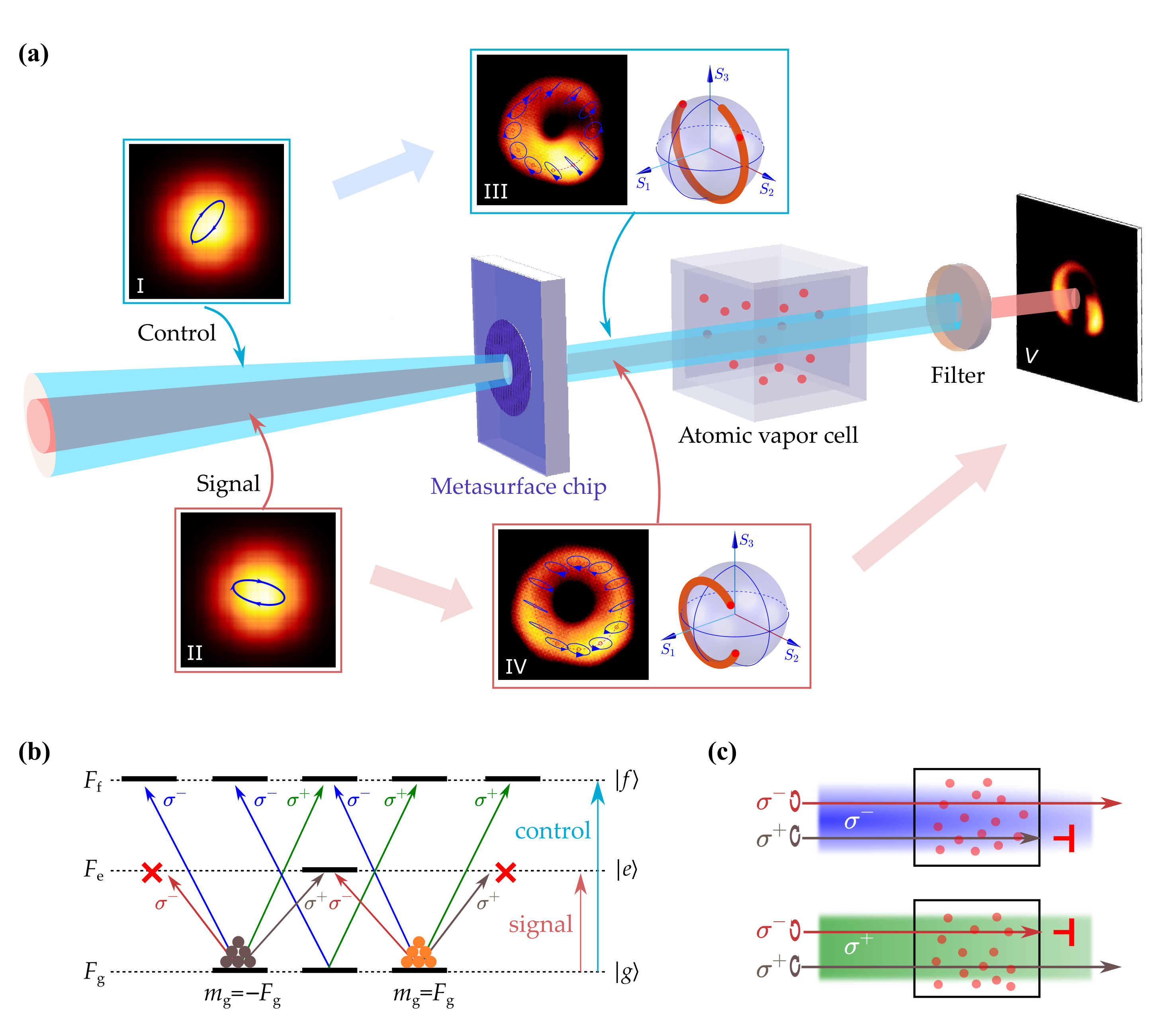}
\par\end{centering}
\caption{(a) The schematic diagram illustrates intensity modulation in vectorial
dual beams produced by combining a single metasurface chip with a
thermal atomic vapor cell. Both the input control light (I) and the
signal light (II) are Gaussian beams, each displaying unique elliptical
polarizations. The corresponding vectorial dual beams (III, IV) generated
by the same metasurface exhibit their polarization states on the Poincar\'{e}
sphere. The intensity profile of the output signal beam (V) is clipped
due to interactions with thermal atoms. (b) The schematic illustrates
the atomic energy levels and transitions facilitated by interactions
with control and signal lights. The atomic transition $\left|g\right\rangle \rightarrow\left|f\right\rangle $
is induced by interaction with $\sigma^{-}(\sigma^{+})$ polarized
control light. This redistributes the atomic population among ground
magnetic states, enabling a forbidden transition $\left|g\right\rangle \rightarrow\left|e\right\rangle $
for $\sigma^{-}(\sigma^{+})$ polarized signal light, resulting in
circular dichroism within the system. (c) The absorption of the signal
beam by atoms is dependent on optical spin when influenced by control
light. Specifically, when the control light is $\sigma^{-}(\sigma^{+})$
polarized, the signal light with $\sigma^{+}(\sigma^{-})$ polarization
is absorbed, whereas the signal light with $\sigma^{-}(\sigma^{+})$
polarization remains transparent. \label{fig1}}
\end{figure*}

\section{Theory}

Vector beams display an inhomogeneous polarization distribution across
their cross-section. Metasurfaces enhance the generation versatility
of vector beams by meticulously designing the geometry and arrangement
of meta-atoms. When an input beam with a polarization state $\left|s_{\mathrm{in}}\right\rangle $
passes through a meta-atom, it can result in an output beam with a
polarization state $\left|s_{\mathrm{out}}\right\rangle $, where
$\left|s_{\mathrm{in}}\right\rangle $$=\mathrm{\mathbb{J}}\left|s_{\mathrm{out}}\right\rangle $.
In this context, $\mathbb{J}$ denotes the Jones matrix associated
with the meta-atom, as defined by \cite{Zhang2018}

\begin{widetext}

\begin{equation}
\mathbb{J}=\mathrm{e}^{\mathrm{i}\psi_{\mathrm{D}}}\left[\begin{matrix}\cos\left(\frac{\psi_{\mathrm{B}}}{2}\right)+\mathrm{i}\sin\left(\frac{\psi_{\mathrm{B}}}{2}\right)\cos(2\psi_{\mathrm{R}}) & \mathrm{i}\sin\left(\frac{\psi_{\mathrm{B}}}{2}\right)\sin(2\psi_{\mathrm{R}})\\
\mathrm{i}\sin\left(\frac{\psi_{\mathrm{B}}}{2}\right)\sin(2\psi_{\mathrm{R}}) & \cos\left(\frac{\psi_{\mathrm{B}}}{2}\right)-\mathrm{i}\sin\left(\frac{\psi_{\mathrm{B}}}{2}\right)\cos(2\psi_{\mathrm{R}})
\end{matrix}\right],\label{eq:JonesMatrix}
\end{equation}

\end{widetext}

\noindent where $\psi_{\mathrm{D}}$ denotes the dynamic phase introduced,
$\psi_{\mathrm{B}}$ signifies the birefringent phase difference between
the two eigen-polarizations, $\psi_{\mathrm{R}}$ represents the orientation
angle of the meta-atom. By fine-tuning the three parameters of the
meta-atom, it is possible to produce various polarization states in
the output light, even when the incident light has a fixed polarization.
Consequently, by arranging the meta-atoms to create spatially varying
distributions of $\left\{ \psi_{\mathrm{D}},\psi_{\mathrm{B}},\psi_{\mathrm{R}}\right\} $,
the transformation from scalar beams to vector beams can be effectively
achieved. Additionally, incorporating the geometric phase into the
transformation process allows for the creation of vector beams that
exhibit distinct polarization profiles corresponding to various input
polarization states. 

As illustrated in \figref{fig1}a, the incident light is an elliptically
polarized Gaussian beam that passes through a metasurface chip. Consequently,
the polarization profile of the resulting vector beams can simultaneously
encompass various elliptically polarized fields. Further modifications
to the input polarization state result in changes to the polarization
profiles of the generated beams, as illustrated by transformations
between \figref{fig1}{a.I,II} to \figref{fig1}{a.III,IV}. The
distribution of polarization states can be represented by the Poincar\'{e}
spheres depicted in \figref{fig1}{a.III,IV}. Utilizing the designed
metasurface, vector beams can be generated and modified by adjusting
the polarization state of the input light. However, for a given metasurface
structure, the output profile remains fixed for each specific polarized
input light. In specific practical scenarios, there may be a need
to modify or dynamically adjust the generated vector beams based on
a particular input light. To address this necessity, we introduce
an atomic ensemble and harness the mediating role of atoms to shape
the signal beam using a control beam.

In \figref{fig1}b, a three-level hyperfine structure of an atom
is illustrated, consisting of the ground state $\left|g\right\rangle $
and two excited states $\left|e\right\rangle $ and $\left|f\right\rangle $.
Each state has several degenerate magnetic states, indicated by $\left|F_{\mathrm{g}},m_{\mathrm{g}}\right\rangle $,
$\left|F_{\mathrm{e}},m_{\mathrm{e}}\right\rangle $, and $\left|F_{\mathrm{f}},m_{\mathrm{f}}\right\rangle $,
respectively. In this notation, $\left\{ F_{\mathrm{g}},F_{\mathrm{e}},F_{\mathrm{f}}\right\} $
signify magnitudes of the total atomic angular momentum, while $\left\{ m_{\mathrm{g}},m_{\mathrm{e}},m_{\mathrm{f}}\right\} $
represent the corresponding magnetic quantum numbers. When light interacts
with an atom, an electron absorbs a photon and transitions from the
ground state to an excited state within the atom. According to the
selection rules, $\left|F_{\mathrm{g}},m_{\mathrm{g}}\right\rangle \rightarrow\left|F_{\mathrm{e,f}},m_{\mathrm{e,f}}=m_{\mathrm{g}}-1\right\rangle $
transition occurs for left-circularly polarized light ($\sigma^{-}$),
and $\left|F_{\mathrm{g}},m_{\mathrm{g}}\right\rangle \rightarrow\left|F_{\mathrm{e,f}},m_{\mathrm{e,f}}=m_{\mathrm{g}}+1\right\rangle $
transition occurs for right-circularly polarized light ($\sigma^{+}$).
A circularly polarized strong light resonant with the transition from
$\left|g\right\rangle $ to $\left|f\right\rangle $, serving as the
control light, induces a redistribution of the population among the
atomic magnetic states within an ensemble. As the system approaches
steady state, the majority of atoms occupy the $\left|F_{\mathrm{g}},m_{\mathrm{g}}=-F_{\mathrm{g}}\right\rangle $
for $\sigma^{-}$ polarized light, and $\left|F_{\mathrm{g}},m_{\mathrm{g}}=F_{\mathrm{g}}\right\rangle $
for $\sigma^{+}$ polarized light. 

At this point, a circularly polarized weak light, designated as the
signal light and nearly resonant with the $\left|g\right\rangle $
$\rightarrow\left|e\right\rangle $ transition, is introduced into
the system. The polarization state of control light modulates the
population distribution among ground magnetic states, consequently
influencing the absorption of signal light by atoms. When the control
light is $\sigma^{+}$ polarized as shown in \figref{fig1}b, nearly
all atoms are found in the $\left|F_{\mathrm{g}},m_{\mathrm{g}}=F_{g}\right\rangle $.
Consequently, the signal light with the same $\sigma^{+}$ polarization
remains transparent due to the lack of available transitions. In contrast,
the $\sigma^{-}$ polarized signal light experiences significant absorption.
Conversely, for $\sigma^{+}$ polarized control light, the effects
on the signal lights are reversed. Consequently, when the control
light interacts with an atomic ensemble, pronounced circular dichroism
is exhibited for weak signal light, as shown in \figref{fig1}c.
This interaction leads to a differential absorption effect, where
the atomic vapor interacts differently with the left-handed and right-handed
circularly polarized components of the vector beams, resulting in
the spatially varying absorption patterns within the beam profiles.
The design of metasurface chips plays a pivotal role in generating
vector beams with tailored polarization states, facilitating tunable
atomic absorption and thereby enhancing the versatility of light field
manipulation. Within this setup, a vectorial control beam can manipulate
another vectorial signal beam, thereby enabling the spatial modulation
of light intensity.

When vector beams passing through metasurface chip and entering in
an atomic ensemble, the electric fields of the signal beam $\mathbf{E}_{\mathrm{s}}$
and the control beam $\mathbf{E}_{\mathrm{c}}$, based on the atomic
circular eigen-responses, are decomposed as in the cylindrical coordinate

\begin{equation}
\begin{array}{cc}
\mathbf{E}_{\mathrm{s}}(r,\phi) & =A_{\mathrm{s}}(r,\phi)\left[a_{-}(r,\phi)\mathbf{\hat{e}}_{-}+\ensuremath{a_{+}(r,\phi)\mathbf{\hat{e}}_{+}}\right]\\
\mathbf{E}_{\mathrm{c}}(r,\phi) & =A_{\mathrm{c}}(r,\phi)\left[b_{-}(r,\phi)\mathbf{\hat{e}}_{-}+\ensuremath{b_{+}(r,\phi)\mathbf{\hat{e}}_{+}}\right]
\end{array},\label{eqE-field}
\end{equation}

\noindent where $A_{\mathrm{s(c)}}$ are the amplitude of signal (control)
beams, and $a_{\pm}(b_{\pm})$ are the corresponding normalized components
in the right (left) circularly polarized bases of $\mathbf{\hat{e}}_{+}$
($\mathbf{\hat{e}}_{-}$). In scenarios where both the control and
signal lights are vector beams traversing an atomic vapor cell, their
interaction exhibits spatial variability due to non-uniform polarization
distributions within the atomic ensemble. In regions where both the
controlling and signal fields demonstrate circular polarization, the
transmissivity of the signal light varies depending on whether their
optical spins are antiparallel or parallel. Significant absorption
of the signal light is observed when the control and signal fields
possess opposite circular polarizations. Conversely, transparency
prevails when they share the same circular polarization. Generally,
when two beams are either elliptically or linearly polarized, they
can be decomposed into left- and right-circularly polarized components.
This decomposition leads to varying degrees of absorption for the
signal light. Consequently, the output intensity of the signal light
strongly depends on the spatial polarization distributions of both
the signal and control beams. As shown in \figref{fig1}{a.V}, the
output intensity of signal light ($\mathit{I}_{\mathrm{out}}$) can
be evaluated by

\begin{equation}
\mathit{I}_{\mathrm{out}}(r,\phi)\propto\mathit{A}_{\mathrm{s}}^{2}\exp\left[-\frac{2\pi\kappa l}{\lambda_{\mathrm{s}}}(1-\mathrm{\mathbf{\mathbf{S}_{\mathrm{s}}\cdot\mathbf{S}_{\mathrm{c}}}})\right],\label{eqOutputIntensity}
\end{equation}

\noindent where $\lambda_{\mathrm{s}}$ is the vacuum wavelength of
signal light, $\kappa$ corresponds to the maximal absorption coefficient
for signal and control fields with opposite circular polarizations,
$l$ is the length of light-atom interaction, $\mathbf{S}_{\mathrm{s}}$
($\mathbf{S}_{\mathrm{c}}$) denotes the average photon spin of signal
(control) light, whose magnitude can be given by $S_{\mathrm{s}}=a_{+}^{2}-a_{-}^{2}$
and $S_{\mathrm{c}}=b_{+}^{2}-b_{-}^{2}$, respectively. The detailed
derivations are provided in the Supplementary Materials.

\section{EXPERIMENT AND RESULT}

\begin{figure}
\begin{centering}
\includegraphics[width=8.5cm]{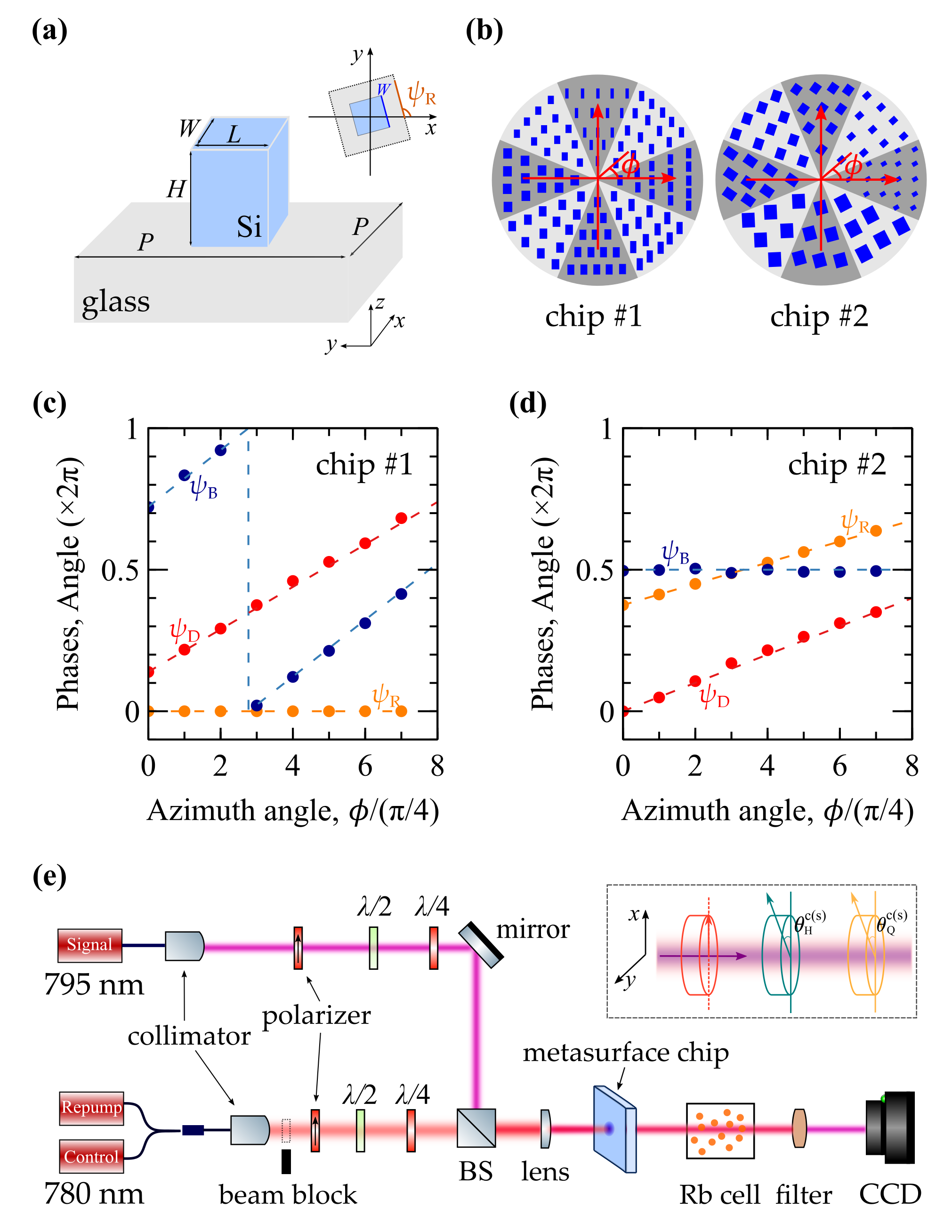}
\par\end{centering}
\caption{(a) Schematic of the metasurface's meta-atom composed of silicon nanofins
on glass substrate, featuring the nanofins with length $L$, width
$W$, height $H$, and period $P$. The inset shows the rotation of
a nanofin at an angle $\psi_{\mathrm{R}}$. (b) Configuration of meta-atom
in two metasurfaces chips (\#1, \#2) to generate different distribution
of vector beam. For each azimuth angle ($\phi$), a unique set of
angles $\left\{ \psi_{\mathrm{D}},\psi_{\mathrm{B}},\psi_{\mathrm{R}}\right\} $
is used in individual sectors for (c) chip \#1 and (d) chip \#2. (e)
Experimental setup with control/repump (red) and signal (pink) light
beams. It includes wave plates for regulating polarized light, with
an inset diagram showing the rotation angle of each wave plate. A
beam splitter (BS) combines the beams, which then pass through a metasurface
chip and a Rb vapor cell. A $795\,\mathrm{nm}$ narrow bandpass filter
is utilized to ensure that solely the signal light is transmitted
to a CCD camera.\label{fig2}}
\end{figure}

In this study, we engineered meta-atoms for a metasurface to modulate
the parameters $\left\{ \psi_{\mathrm{D}},\psi_{\mathrm{B}},\psi_{\mathrm{R}}\right\} $
within the Jones matrix \cite{Zhang2018,cui2024exploitingcombineddynamicgeometric}.
By integrating the geometric phase effect, we generated vector beams
exhibiting diverse linear and elliptical polarization profiles, tailored
to various polarization states of input light. As shown in \figref{fig2}a,
the metasurface pattern was fabricated on a $1\text{-}\mathrm{mm}$
thick silicon-on-glass substrate and arranged in a circular configuration
with a diameter of $0.5\,\mathrm{mm}$. To achieve spatially varied
polarization profiles for the output beams, we segmented the metasurface
pattern into eight fan-shaped sectors. Each sector contains meta-atoms
of a specific design, all sharing the same dimensions. Meta-atoms
are constructed from silicon nanofins, each with a height ($H$) of
$400\,\mathrm{nm}$, dimensions that vary in length ($\mathrm{\mathit{L}}$)
and width ($\mathrm{\mathit{W}}$) from $110\,\mathrm{nm}$ to $240\,\mathrm{nm}$.
These nanofins are arranged in a square array with a pitch ($P$)
of $400\,\mathrm{nm}$ and oriented at angles $\psi_{\mathrm{R}}$
ranging from $0^{\circ}$ to $360^{\circ}$ according to design requirements\textcolor{black}{.
}In the experiment, two distinct metasurfaces were meticulously designed
as illustrated in \figref{fig2}b. By employing a variety of nanofin
configurations, it was possible to achieve different sets of $\left\{ \psi_{\mathrm{D}},\psi_{\mathrm{B}},\psi_{\mathrm{R}}\right\} $,
thereby facilitating the generation of vector beams with diverse polarization
distributions. In \figsref{fig2}{c-d}, the symbols at each azimuth
angle ($\phi$) represent a unique configuration $\left\{ \psi_{\mathrm{D}},\psi_{\mathrm{B}},\psi_{\mathrm{R}}\right\} $
employed for an individual sector. The eight sectors incorporate diverse
silicon nanofin dimensions and arrangements\textcolor{black}{{} (see
the Supplementary Materials)}, generating vector beams with various
polarization profiles for different input light polarization states.

In the experiment, an atomic ensemble was realized using a vapor cell
containing rubidium (Rb) atoms. Referring to the level structure shown
in \figref{fig1}b, we employed $^{87}\mathrm{Rb}$ atoms and selected
the $F_{\mathrm{g}}=2\rightarrow F_{\mathrm{f}}=3$ transition of
the D2 line as the $\left|g\right\rangle $ $\rightarrow\left|f\right\rangle $
transition, as well as the $F_{\mathrm{g}}=2\rightarrow F_{\mathrm{e}}=1$
transition of the D1 line as the $\left|g\right\rangle $ $\rightarrow\left|e\right\rangle $
transition\textcolor{black}{{} (see the Supplementary Materials)}. In
practical scenarios, when atoms transition from excited states back
to the ground state, some of them might undergo de-excitation to the
$5^{2}S_{1/2}(F=1)$ energy level, thereby exiting the three-level
transition cycles under consideration. To address this issue, we introduced
a repump light that resonates with the $F=1\rightarrow F_{\mathrm{f}}=3$
transition of the D2 line. This ensures that the atoms are re-excited
and returned to the relevant transition cycles.

\begin{figure}
\begin{centering}
\includegraphics[width=8cm]{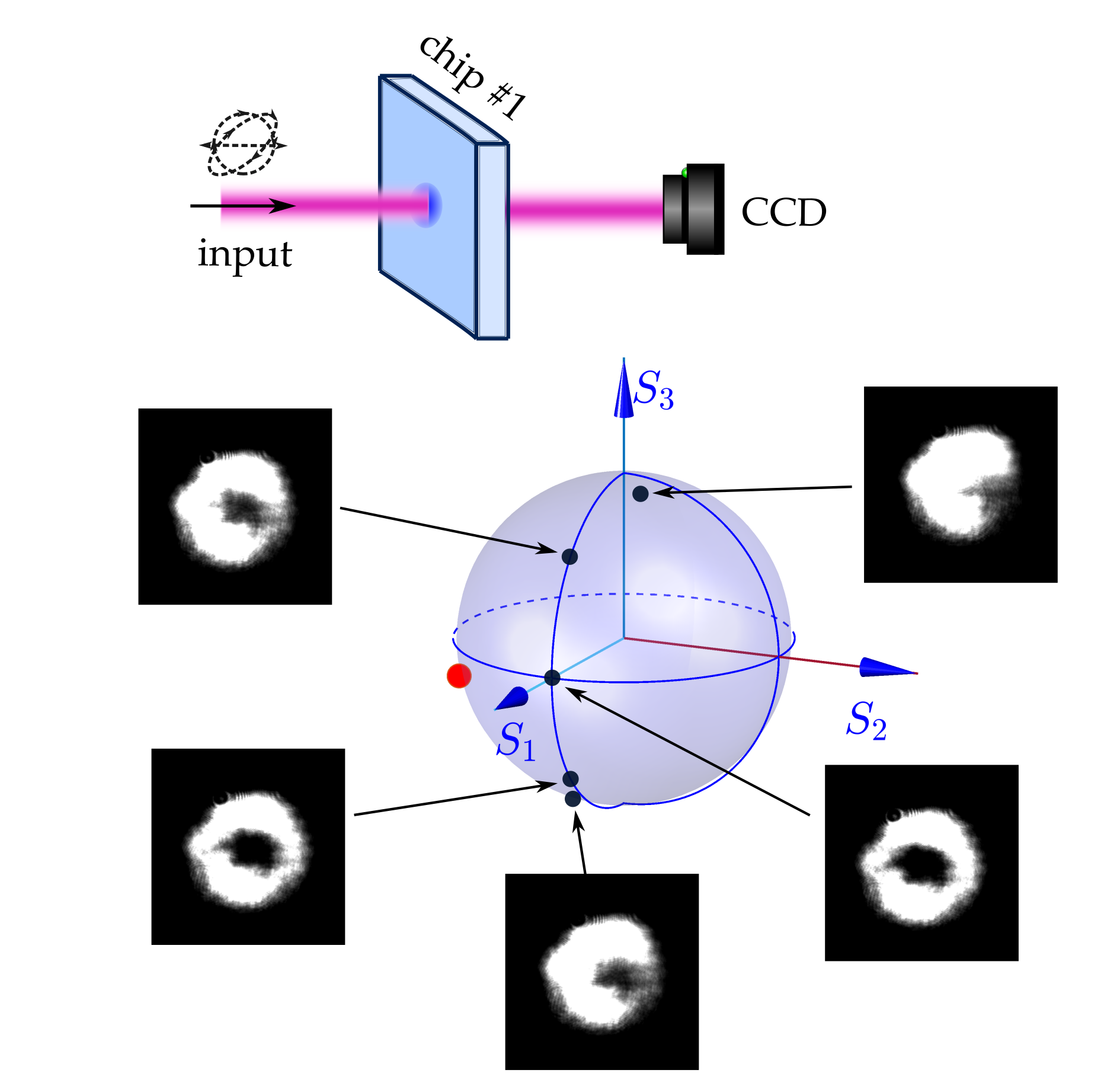}
\par\end{centering}
\caption{Intensity distribution of the signal vector beam generated by metasurface
chip \#1 varies with changes in the polarization state of the incident
light, as represented on a Poincar\'{e} sphere.\label{fig3}}
\end{figure}

\begin{figure*}
\begin{centering}
\includegraphics[width=14cm]{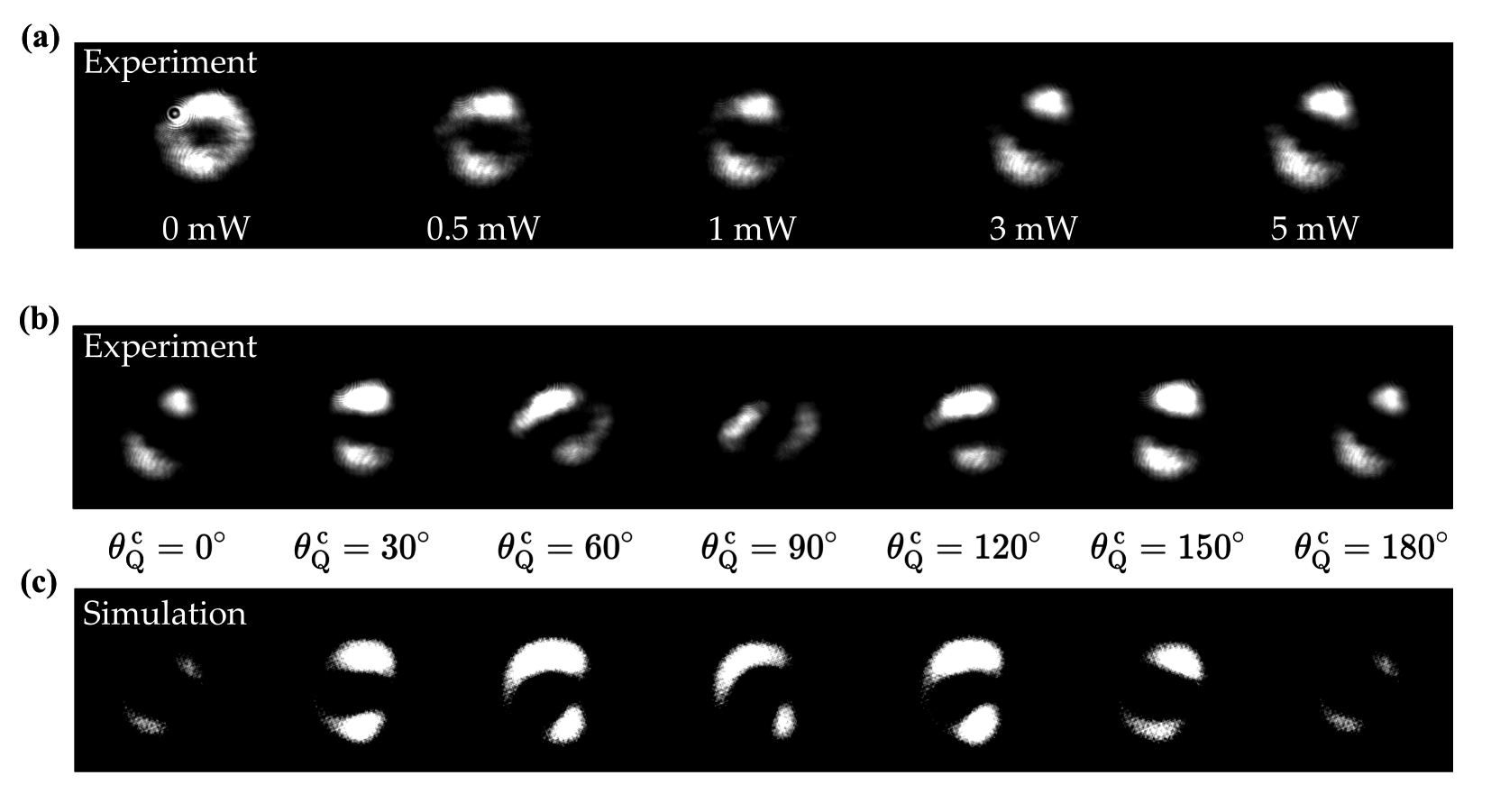}
\par\end{centering}
\caption{Experimental and simulation results of the intensity distribution
for the signal vector beam modulated by the control vector beam, generated
using chip \#1. (a) Intensity distribution of the signal vector beam
while varying the power of the control incident light from $0\,\mathrm{mW}$
to $5\,\mathrm{mW}$. (b) The experimental and (c) simulated results
for the intensity distribution of the signal beam with $\theta_{\mathrm{Q}}^{\mathrm{c}}$
on the control light varying from $0^{\circ}$to $180^{\circ}$, respectively.
In the experiment, the powers of signal, control and repump lights
are set as $0.04\,\mathrm{mW}$, $5.5\,\mathrm{mW}$ and $1\,\mathrm{mW}$,
respectively. \label{fig4}}
\end{figure*}

The experimental system, as illustrated in \figref{fig2}e, was
constructed based on a metasurface chip and an atomic Rb vapor cell
to demonstrate the modulation of vector beams. The system includes
a $795\,\mathrm{nm}$ laser for the signal light and two $780\,\mathrm{nm}$
lasers serving as the control and repump lights. The control and repump
lights are merged into a single path using a $50:50$ fiber coupler,
subsequently converted into a free-space beam via a fiber collimator.
The polarization state of the incident light is tuned by a combination
of a polarizer, half-wave ($\lambda/2$) plate, and quarter-wave ($\lambda/4$)
plate. As illustrated in the inset of \figref{fig2}e, the signal
(control) light is initially polarized along the $x$-axis by the
polarizer. Subsequently, to alter the polarization states, the $\lambda/2$
plate is aligned at angles $\theta_{\mathrm{H}}^{\mathrm{s}}$ ($\theta_{\mathrm{H}}^{\mathrm{c}}$)
and the $\lambda/4$ plate is positioned at angles $\theta_{\mathrm{Q}}^{\mathrm{s}}$
($\theta_{\mathrm{Q}}^{\mathrm{c}})$ relative to the polarizer. The
signal light is combined with the control light via a beam splitter.
Upon optimizing the spot size via a lens, the signal, control, and
repump lights are concurrently transmitted across the metasurface
chip. By adjusting the incident laser power, we facilitate the population
of atoms in their ground state by the control and repump lights, meticulously
avoiding saturation absorption. In the experiment, The side length
of the cubic vapor cell is $20\,\mathrm{mm}$, and to enhance the
atomic vapor density, the temperature of the cell was maintained at
$75\mathrm{{^\circ}C}$. At the endpoint of the optical paths, a $795\,\mathrm{nm}$
narrow bandpass filter is implemented to block $780\,\mathrm{nm}$
lights, thereby precluding interference from control and repump beams
on the final output. As a result, the signal light proceeds to enter
a CCD camera to acquire intensity image of the signal beam.

With the specifically designed metasurface chip, the shape of the
light beam can be adjusted by tuning the polarization of the input
light. As illustrated in the inset of \figref{fig3}{}, the Gaussian-distributed
signal beam passes through chip \#1 with varying polarization states
(without control light). The output intensity patterns are illustrated
in \figref{fig3}{} for the respective input polarization states
indicated on a Poincar\'{e} sphere. The capability to modulate the
intensity distribution of the output signal beam is effectively demonstrated
through the sole adjustment of the input signal light's polarization.
However, the output beams are in forms of doughnut or crescent shapes,
even for all the other input polarization states. Herein, we demonstrate
that the signal beams can be further tailored by incorporating a control
beam. To elucidate this manipulation, we maintain a constant power
of $40\,\mu\mathrm{W}$ and a specific polarization state of the input
signal light, as depicted by a red circular marker on the Poincar\'{e}
sphere in \figref{fig3}{}. This configuration yields a doughnut-shaped
output intensity distribution when utilizing chip \#1. Subsequently,
the control light is introduced to further tailor the signal beam's
profile. Figure \fref{fig4}a illustrates the intensity distribution
of the signal beam as a function of the varying incident power from
$0\,\mathrm{mW}$ to $5\,\mathrm{mW}$ of the control light, with
its polarization states set by $\{\theta_{\mathrm{H}}^{\mathrm{c}},\theta_{\mathrm{Q}}^{\mathrm{c}}\}=\{20^{\circ},0^{\circ}\}$.
Upon introducing the control light into the system and gradually increasing
its incident power, the doughnut-shaped signal beam bifurcates into
two lobes due to inhomogeneous absorption induced by the atomic vapor.
The observed inhomogeneity is attributed to the circular dichroism
that arises from the interaction of atoms with the distinct vector
beams of signal and control light, which are generated by the metasurface
chip.

Indeed, adjusting the incident polarization of the control beam permits
enhanced flexibility in sculpting the intensity profile of the signal
beam. In the experiment, the power of the control light was set at
$5.5\,\mathrm{mW}$, with its polarization settings configured as
$\theta_{\mathrm{H}}^{\mathrm{c}}=20^{\circ}$ and $\theta_{\mathrm{Q}}^{\mathrm{c}}$
varied from $0^{\circ}$ to $180^{\circ}$. The intensity distribution
of the modulated signal beam is depicted in \figref{fig4}b, where
the two distinct lobed shapes rotate due to varying polarizations
of the control light. The intensity pattern of the signal beam rotates
counterclockwise as $\theta_{\mathrm{Q}}^{\mathrm{c}}$ varies from
$0^{\circ}$ to $90^{\circ}$ and clockwise as it ranges from $90^{\circ}$
to $180^{\circ}$. The observed variation is directly attributable
to the control vector beam's modulation of atomic absorption across
various incident polarization states. The simulated results, obtained
using equation (\ref{eqOutputIntensity}), are illustrated in \figref{fig4}c
and exhibit a satisfactory correspondence with the experimental findings.

\begin{figure}
\begin{centering}
\includegraphics[width=8cm]{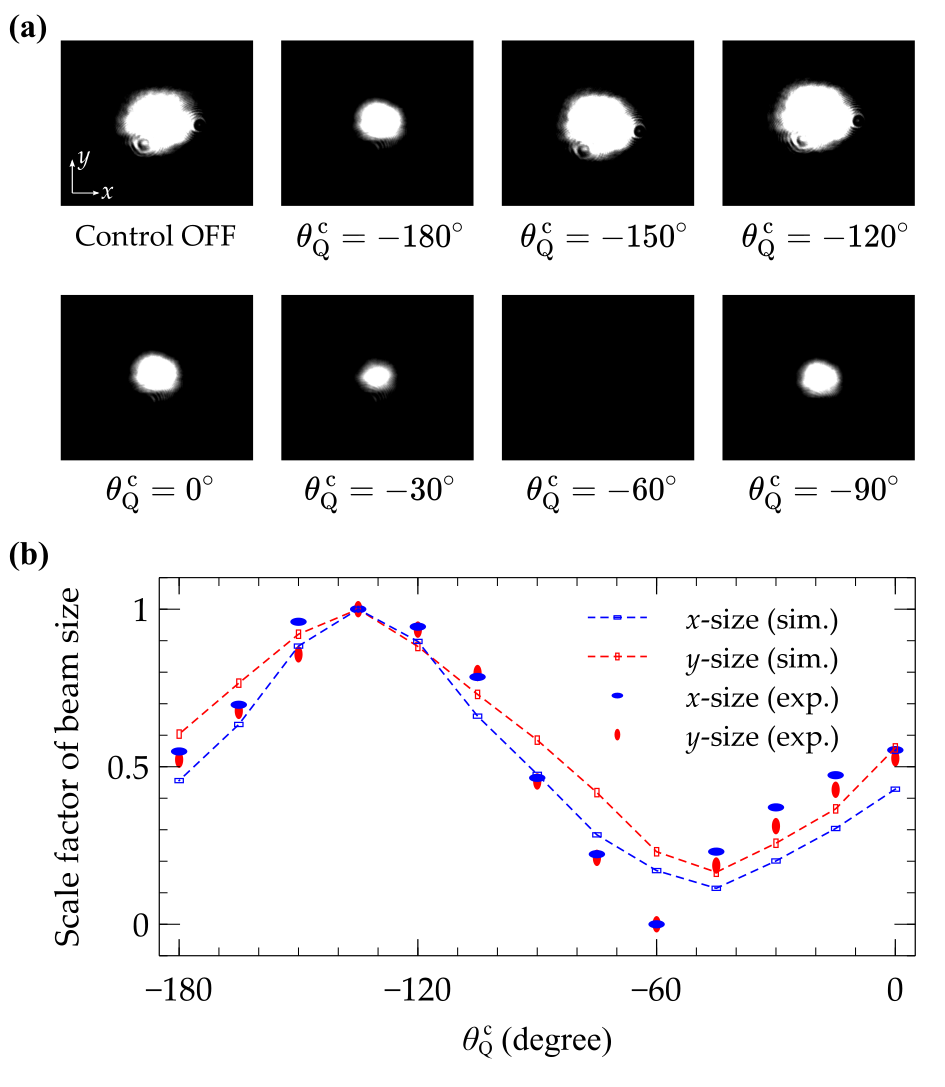}
\par\end{centering}
\caption{(a) Experimental results of intensity distribution for signal beam
modulated by control light, with $\theta_{\mathrm{H}}^{\mathrm{c}}=0^{\circ}$
and $\theta_{\mathrm{Q}}^{\mathrm{c}}$ varying from $0^{\circ}$
to $-180^{\circ}$, under chip \#2. (b) Experimental and simulation
results of scale factor of beam size versus $\theta_{\mathrm{Q}}^{\mathrm{c}}$
for the control light. The blue (red) ellipse symbol represents the
beam size along the $x$-axis ($y$-axis) in the experimental results,
while the blue (red) dashed line with a bar signifies the beam size
along the $x$-axis ($y$-axis) in the simulation results.\label{fig5}}
\end{figure}

Moreover, utilizing chip \#2 and fine-tuning the polarization state
of the incoming signal light, we produce a beam whose intensity distribution
closely approximates a Gaussian profile, as depicted in the first
pattern of \figref{fig5}a. Upon activation of the control light,
with $\theta_{\mathrm{H}}^{\mathrm{c}}=0^{\circ}$ and $\theta_{\mathrm{Q}}^{\mathrm{c}}$
varying from $0^{\circ}$ to $-180^{\circ}$, the beam size of the
Gaussian-distributed signal beams can be adjusted in response to changes
in the control polarization state. Figure \fref{fig4}b presents
the scale factor of the beam size along both the $x$- and $y$-axes,
comparing experimental data with calculated results (also see the
Supplementary materials). It is clear that as $\theta_{\mathrm{Q}}^{\mathrm{c}}$
of the control light varies from $0^{\circ}$ to $-180^{\circ}$,
the beam size can be altered by nearly an order of magnitude. This
suggests that the shape of the signal beam is significantly influenced
and tailored by the control vector beam through the thermal atomic
vapor system. The discrepancies observed between the experimental
and simulated results of $\theta_{\mathrm{Q}}^{\mathrm{c}}$ ranging
from $-60^{\circ}$ to $0^{\circ}$ are primarily attributed to deviations
from the design specifications due to fabrication imperfections. The
absence of a visible spot for $\theta_{\mathrm{Q}}^{\mathrm{c}}=-60^{\circ}$
as depicted in \figref{fig5}a is a consequence of the fixed exposure
settings of the CCD camera.

\section{Discussion }

This study presents a system designed for vectorial dual-beam interactions
with thermal atoms, specifically engineered for intensity clipping
of vector beams. We utilize metasurface chips to generate dual vectorial
beams: one as the control vector beam and the other as the signal
vector beam. The vectorial dual-beam clipping method modulates the
intensity of the signal vector beam without the need to adjust the
parameters of the signal light itself. This method enhances both the
robustness of the detection optical path and the modulation capabilities
of vector beams.

In this study, we have developed two unique metasurface chip designs
to demonstrate the generation and modulation of vector beams. However,
the vector beams generated are not confined to these two designs alone.
By precisely engineering metasurfaces with specific configurations
of meta-atoms, a diverse distribution of light field outputs can be
achieved. This flexibility empowers a multitude of applications through
vector beam modulation, such as optical manipulations of micro-nano
particles. Our experimental results reveal that the intensity distribution
of the signal beam, when modulated by a control vector beam passing
through two distinct metasurface chips, exhibits two characteristic
patterns: a dual-lobed configuration and a Gaussian-like profile.
The modulation of the polarization state of the control light facilitates
the rotation of the dual lobes and the dynamic adjustment of the dimensions
of the Gaussian-like profile, which in turn facilitates the efficient
trapping and manipulation of micro-nano particles, even at the atomic
level. Thus, this dynamic modulation technique possesses considerable
potential for applications in atomic trapping, transportation, and
associated fields. Furthermore, it also presents promising prospects
in the realms of image processing and quantum information.

The metasurface is fabricated using micro-nano technologies; however,
unavoidable manufacturing imperfections can introduce discrepancies
between the actual structure and its intended design. As a result,
these deviations may cause differences between the actual performance
of the generated vector beams and theoretical predictions. Our simulations
are predicated on ideal design parameters, which can lead to minor
discrepancies between the theoretical predictions and the actual experimental
results. In our simulations, the results are obtained using equation
(\ref{eqOutputIntensity}) which is derived from a phenomenological
model (further details are provided in the Supplementary Materials).
However, this simple model inaccurately captures the intermediate
states between two perfect circular polarizations, leading to less
accurate predictions for these scenarios. The precise methodology
outlined in the Supplementary Materials is derived from the general
atom-light interaction process. Nevertheless, determining the spatially
variant atomic-induced susceptibility for vector beams persists as
a significant challenge, primarily due to the voluminous nature of
the required computations.

\section{Conclusion}

This study introduces a novel hybrid system that combines metasurfaces
with thermal atomic vapor to precisely modulate vector beam intensity
profiles. A vectorial dual-beam configuration is introduced, where
both the control and signal beams are generated using a single metasurface
chip. The inherent flexibility in the design of metasurfaces enables
the generation of distinct vector beams through adjustment of their
meta-atom arrangements and geometric configurations. By utilizing
the absorption characteristics of atomic vapor, a medium with optical
spin-dependent absorption is constructed, stemming from the control
light modifying the atomic population distribution. The circular dichroism
feature is utilized to effectively modulate the intensity distribution
of the signal beam using the control beam. To demonstrate diverse
modulation, we designed and fabricated two metasurface chips capable
of generating vector beams with various polarization distributions.
By adjusting the power and polarization states of control light, we
successfully tailored the intensity profiles of both doughnut-shaped
and Gaussian-distributed signal beams. These advancements hold significant
application prospects in particle trapping and manipulation, image
processing, and quantum information.

\bigskip{}

\noindent \textbf{Funding}

\noindent This work is supported by the Fundamental Research Funds
for the Central Universities under Grant KG21008401.

\medskip{}

\noindent \textbf{Disclosures. }The authors declare no competing financial
interests.

\noindent \textbf{Data availability. }Data underlying the results
presented in this paper are not publicly available at this time but
may be obtained from the authors upon reasonable request.

\medskip{}

\noindent \textbf{Supporting Information. }Detailed theoretical derivation
of the effective electric susceptibility of atomic vapor, experimental
results of beam shaping, geometric parameters of meta-atoms, and simulated
polarization distributions of vector beams.

\bibliographystyle{naturemag}

\begin{thebibliography}{10}
\expandafter\ifx\csname url\endcsname\relax
  \def\url#1{\texttt{#1}}\fi
\expandafter\ifx\csname urlprefix\endcsname\relax\def\urlprefix{URL }\fi
\providecommand{\bibinfo}[2]{#2}
\providecommand{\eprint}[2][]{\url{#2}}

\bibitem{Pattanayak1980}
\bibinfo{author}{Pattanayak, D.} \& \bibinfo{author}{Agrawal, G.}
\newblock \bibinfo{title}{Representation of vector electromagnetic beams}.
\newblock \emph{\bibinfo{journal}{Phys. Rev. A}} \textbf{\bibinfo{volume}{22}},
  \bibinfo{pages}{1159} (\bibinfo{year}{1980}).

\bibitem{Hall1996}
\bibinfo{author}{Hall, D.~G.}
\newblock \bibinfo{title}{Vector-beam solutions of {{M}}axwell's wave
  equation}.
\newblock \emph{\bibinfo{journal}{Opt. Lett.}} \textbf{\bibinfo{volume}{21}},
  \bibinfo{pages}{9--11} (\bibinfo{year}{1996}).

\bibitem{matailoring}
\bibinfo{author}{Ma, J.}, \bibinfo{author}{Xie, Z.} \& \bibinfo{author}{Yuan,
  X.}
\newblock \bibinfo{title}{Tailoring arrays of optical stokes skyrmions in
  tightly focused beams}.
\newblock \emph{\bibinfo{journal}{Laser Photonics Rev.}}
  \bibinfo{pages}{2401113} (\bibinfo{year}{2024}).

\bibitem{Kumar2024}
\bibinfo{author}{Kumar, R.~N.}, \bibinfo{author}{Nayak, J.~K.},
  \bibinfo{author}{Gupta, S.~D.}, \bibinfo{author}{Ghosh, N.} \&
  \bibinfo{author}{Banerjee, A.}
\newblock \bibinfo{title}{Probing dual asymmetric transverse spin angular
  momentum in tightly focused vector beams in optical tweezers}.
\newblock \emph{\bibinfo{journal}{Laser Photonics Rev.}}
  \textbf{\bibinfo{volume}{18}}, \bibinfo{pages}{2300189}
  (\bibinfo{year}{2024}).

\bibitem{donato2012optical}
\bibinfo{author}{Donato, M.} \emph{et~al.}
\newblock \bibinfo{title}{Optical trapping of nanotubes with cylindrical vector
  beams}.
\newblock \emph{\bibinfo{journal}{Opt. Lett.}} \textbf{\bibinfo{volume}{37}},
  \bibinfo{pages}{3381--3383} (\bibinfo{year}{2012}).

\bibitem{kozawa2010optical}
\bibinfo{author}{Kozawa, Y.} \& \bibinfo{author}{Sato, S.}
\newblock \bibinfo{title}{Optical trapping of micrometer-sized dielectric
  particles by cylindrical vector beams}.
\newblock \emph{\bibinfo{journal}{Opt. Express}} \textbf{\bibinfo{volume}{18}},
  \bibinfo{pages}{10828--10833} (\bibinfo{year}{2010}).

\bibitem{Milione2015}
\bibinfo{author}{Milione, G.}, \bibinfo{author}{Nguyen, T.~A.},
  \bibinfo{author}{Leach, J.}, \bibinfo{author}{Nolan, D.~A.} \&
  \bibinfo{author}{Alfano, R.~R.}
\newblock \bibinfo{title}{Using the nonseparability of vector beams to encode
  information for optical communication}.
\newblock \emph{\bibinfo{journal}{Opt. Lett.}} \textbf{\bibinfo{volume}{40}},
  \bibinfo{pages}{4887--4890} (\bibinfo{year}{2015}).

\bibitem{Zhu2021}
\bibinfo{author}{Zhu, Z.} \emph{et~al.}
\newblock \bibinfo{title}{Compensation-free high-dimensional free-space optical
  communication using turbulence-resilient vector beams}.
\newblock \emph{\bibinfo{journal}{Nat. Commun.}} \textbf{\bibinfo{volume}{12}},
  \bibinfo{pages}{1666} (\bibinfo{year}{2021}).

\bibitem{zhao2015high}
\bibinfo{author}{Zhao, Y.} \& \bibinfo{author}{Wang, J.}
\newblock \bibinfo{title}{High-base vector beam encoding/decoding for
  visible-light communications}.
\newblock \emph{\bibinfo{journal}{Opt. Lett.}} \textbf{\bibinfo{volume}{40}},
  \bibinfo{pages}{4843--4846} (\bibinfo{year}{2015}).

\bibitem{Syubaev2019}
\bibinfo{author}{Syubaev, S.} \emph{et~al.}
\newblock \bibinfo{title}{Plasmonic nanolenses produced by cylindrical vector
  beam printing for sensing applications}.
\newblock \emph{\bibinfo{journal}{Sci. Rep.}} \textbf{\bibinfo{volume}{9}},
  \bibinfo{pages}{19750} (\bibinfo{year}{2019}).

\bibitem{wu2021cylindrical}
\bibinfo{author}{Wu, H.} \emph{et~al.}
\newblock \bibinfo{title}{Cylindrical vector beam for vector magnetic field
  sensing based on magnetic fluid}.
\newblock \emph{\bibinfo{journal}{IEEE Photonics Technol. Lett.}}
  \textbf{\bibinfo{volume}{33}}, \bibinfo{pages}{703--706}
  (\bibinfo{year}{2021}).

\bibitem{Guo2017}
\bibinfo{author}{Guo, Q.} \emph{et~al.}
\newblock \bibinfo{title}{Manipulation of vector beam polarization with
  geometric metasurfaces}.
\newblock \emph{\bibinfo{journal}{Opt. Express}} \textbf{\bibinfo{volume}{25}},
  \bibinfo{pages}{14300--14307} (\bibinfo{year}{2017}).

\bibitem{wen2022broadband}
\bibinfo{author}{Wen, D.} \emph{et~al.}
\newblock \bibinfo{title}{Broadband multichannel cylindrical vector beam
  generation by a single metasurface}.
\newblock \emph{\bibinfo{journal}{Laser Photonics Rev.}}
  \textbf{\bibinfo{volume}{16}}, \bibinfo{pages}{2200206}
  (\bibinfo{year}{2022}).

\bibitem{zhang2018optical}
\bibinfo{author}{Zhang, C.} \emph{et~al.}
\newblock \bibinfo{title}{Optical metasurface generated vector beam for
  anticounterfeiting}.
\newblock \emph{\bibinfo{journal}{Phys. Rev. Appl.}}
  \textbf{\bibinfo{volume}{10}}, \bibinfo{pages}{034028}
  (\bibinfo{year}{2018}).

\bibitem{fu2023metasurface}
\bibinfo{author}{Fu, P.} \emph{et~al.}
\newblock \bibinfo{title}{Metasurface enabled on-chip generation and
  manipulation of vector beams from vertical cavity surface-emitting lasers}.
\newblock \emph{\bibinfo{journal}{Adv. Mater.}} \textbf{\bibinfo{volume}{35}},
  \bibinfo{pages}{2204286} (\bibinfo{year}{2023}).

\bibitem{cui2024multifunctional}
\bibinfo{author}{Cui, G.} \emph{et~al.}
\newblock \bibinfo{title}{Multifunctional all-dielectric quarter-wave plate
  metasurfaces for generating focused vector beams of bell-like states}.
\newblock \emph{\bibinfo{journal}{Proc. Spie.}} \textbf{\bibinfo{volume}{13}},
  \bibinfo{pages}{1631--1644} (\bibinfo{year}{2024}).

\bibitem{zhu2022multidimensional}
\bibinfo{author}{Zhu, J.-L.} \emph{et~al.}
\newblock \bibinfo{title}{Multidimensional trapping by dual-focusing
  cylindrical vector beams with all-silicon metalens}.
\newblock \emph{\bibinfo{journal}{Photonics Res.}}
  \textbf{\bibinfo{volume}{10}}, \bibinfo{pages}{1162--1169}
  (\bibinfo{year}{2022}).

\bibitem{Kim2021}
\bibinfo{author}{Kim, I.} \emph{et~al.}
\newblock \bibinfo{title}{Pixelated bifunctional metasurface-driven dynamic
  vectorial holographic color prints for photonic security platform}.
\newblock \emph{\bibinfo{journal}{Nat. Commun.}} \textbf{\bibinfo{volume}{12}},
  \bibinfo{pages}{3614} (\bibinfo{year}{2021}).

\bibitem{arbabi2015dielectric}
\bibinfo{author}{Arbabi, A.}, \bibinfo{author}{Horie, Y.},
  \bibinfo{author}{Bagheri, M.} \& \bibinfo{author}{Faraon, A.}
\newblock \bibinfo{title}{Dielectric metasurfaces for complete control of phase
  and polarization with subwavelength spatial resolution and high
  transmission}.
\newblock \emph{\bibinfo{journal}{Nat. Nanotechnol.}}
  \textbf{\bibinfo{volume}{10}}, \bibinfo{pages}{937--943}
  (\bibinfo{year}{2015}).

\bibitem{deng2018diatomic}
\bibinfo{author}{Deng, Z.-L.} \emph{et~al.}
\newblock \bibinfo{title}{Diatomic metasurface for vectorial holography}.
\newblock \emph{\bibinfo{journal}{Nano Lett.}} \textbf{\bibinfo{volume}{18}},
  \bibinfo{pages}{2885--2892} (\bibinfo{year}{2018}).

\bibitem{Cai2024}
\bibinfo{author}{Cai, G.}, \bibinfo{author}{Tian, K.} \& \bibinfo{author}{Wang,
  Z.}
\newblock \bibinfo{title}{Thermal atomic compass based on radially polarized
  beam}.
\newblock \emph{\bibinfo{journal}{Laser Photonics Rev.}}
  \bibinfo{pages}{2400465} (\bibinfo{year}{2024}).

\bibitem{Castellucci2021}
\bibinfo{author}{Castellucci, F.}, \bibinfo{author}{Clark, T.~W.},
  \bibinfo{author}{Selyem, A.}, \bibinfo{author}{Wang, J.} \&
  \bibinfo{author}{Franke-Arnold, S.}
\newblock \bibinfo{title}{Atomic compass: detecting 3{{D}} magnetic field
  alignment with vector vortex light}.
\newblock \emph{\bibinfo{journal}{Phys. Rev. Lett.}}
  \textbf{\bibinfo{volume}{127}}, \bibinfo{pages}{233202}
  (\bibinfo{year}{2021}).

\bibitem{Qiu2021}
\bibinfo{author}{Qiu, S.} \emph{et~al.}
\newblock \bibinfo{title}{Visualization of magnetic fields with cylindrical
  vector beams in a warm atomic vapor}.
\newblock \emph{\bibinfo{journal}{Photonics Res.}}
  \textbf{\bibinfo{volume}{9}}, \bibinfo{pages}{2325--2331}
  (\bibinfo{year}{2021}).

\bibitem{Chang2023}
\bibinfo{author}{Chang, H.} \emph{et~al.}
\newblock \bibinfo{title}{Atomic optical spatial mode extractor for vector
  beams based on polarization-dependent absorption}.
\newblock \emph{\bibinfo{journal}{Chin. Phys. B}}
  \textbf{\bibinfo{volume}{32}}, \bibinfo{pages}{034207}
  (\bibinfo{year}{2023}).

\bibitem{Ye2019}
\bibinfo{author}{Ye, Y.-H.}, \bibinfo{author}{Dong, M.-X.},
  \bibinfo{author}{Yu, Y.-C.}, \bibinfo{author}{Ding, D.-S.} \&
  \bibinfo{author}{Shi, B.-S.}
\newblock \bibinfo{title}{Experimental realization of optical storage of vector
  beams of light in warm atomic vapor}.
\newblock \emph{\bibinfo{journal}{Opt. Lett.}} \textbf{\bibinfo{volume}{44}},
  \bibinfo{pages}{1528--1531} (\bibinfo{year}{2019}).

\bibitem{Yang2019}
\bibinfo{author}{Yang, X.} \emph{et~al.}
\newblock \bibinfo{title}{Observing quantum coherence induced transparency of
  hybrid vector beams in atomic vapor}.
\newblock \emph{\bibinfo{journal}{Opt. Lett.}} \textbf{\bibinfo{volume}{44}},
  \bibinfo{pages}{2911--2914} (\bibinfo{year}{2019}).

\bibitem{Sun2023}
\bibinfo{author}{Sun, Y.} \& \bibinfo{author}{Wang, Z.}
\newblock \bibinfo{title}{Optically polarized selective transmission of a
  fractional vector vortex beam by the polarized atoms with external magnetic
  fields}.
\newblock \emph{\bibinfo{journal}{Opt. Express}} \textbf{\bibinfo{volume}{31}},
  \bibinfo{pages}{15409--15422} (\bibinfo{year}{2023}).

\bibitem{wang2024measuring}
\bibinfo{author}{Wang, J.} \emph{et~al.}
\newblock \bibinfo{title}{Measuring the optical concurrence of vector beams
  with an atomic-state interferometer}.
\newblock \emph{\bibinfo{journal}{Phys. Rev. Lett.}}
  \textbf{\bibinfo{volume}{132}}, \bibinfo{pages}{193803}
  (\bibinfo{year}{2024}).

\bibitem{Sedlacek2012a}
\bibinfo{author}{Sedlacek, J.~A.} \emph{et~al.}
\newblock \bibinfo{title}{Microwave electrometry with rydberg atoms in a vapour
  cell using bright atomic resonances}.
\newblock \emph{\bibinfo{journal}{Nat. Phys.}} \textbf{\bibinfo{volume}{8}},
  \bibinfo{pages}{819--824} (\bibinfo{year}{2012}).

\bibitem{Hu2021}
\bibinfo{author}{Hu, X.-X.} \emph{et~al.}
\newblock \bibinfo{title}{Noiseless photonic non-reciprocity via
  optically-induced magnetization}.
\newblock \emph{\bibinfo{journal}{Nat. Commun.}} \textbf{\bibinfo{volume}{12}},
  \bibinfo{pages}{2389} (\bibinfo{year}{2021}).

\bibitem{Zhang2018}
\bibinfo{author}{Zhang, D.}, \bibinfo{author}{Feng, X.} \&
  \bibinfo{author}{Huang, Y.}
\newblock \bibinfo{title}{Orbital angular momentum induced by nonabsorbing
  optical elements through space-variant polarization-state manipulations}.
\newblock \emph{\bibinfo{journal}{Phys. Rev. A}} \textbf{\bibinfo{volume}{98}},
  \bibinfo{pages}{043845} (\bibinfo{year}{2018}).

\bibitem{cui2024exploitingcombineddynamicgeometric}
\bibinfo{author}{Cui, J.}, \bibinfo{author}{Qing, C.}, \bibinfo{author}{Feng,
  L.} \& \bibinfo{author}{Zhang, D.}
\newblock \bibinfo{title}{Exploiting the combined dynamic and geometric phases
  for optical vortex beam generation using metasurfaces}.
\newblock \emph{\bibinfo{journal}{arXiv}} \bibinfo{pages}{2412.05121}
  (\bibinfo{year}{2024}).

\end{thebibliography}

\end{document}